# Die Einflüsse von Arbeitsbelastung auf die Arbeitsqualität agiler Software-Entwicklungsteams

**Autor*innen:** Christian Sanden(1)*, Kira Karnowski(1)*, Marvin Steinke(1), Michael Neumann(1)* und Lukas Linke(2)

1: Hochschule Hannover, Fakultät IV, Abt. Wirtschaftsinformatik, Ricklinger Stadtweg 120, 30459 Hannover

2: Otto GmbH & Co KG., E-Commerce Solutions & Technology, Werner-Otto-Straße 1-7, 22179 Hamburg

*korrespondierende Autoren:

christian.sanden@stud.hs-hannover.de

kira.karnowski@stud.hs-hannover.de

michael.neumann@hs-hannover.de

**Zusammenfassung:** Durch die Covid 19 Pandemie und der damit einhergehenden Effekte auf die Arbeitswelt ist die Belastung der Mitarbeitenden in einen stärkeren Fokus gerückt worden. Dieser Umstand trifft unter anderem durch den umfassenden Wechsel in die Remote Work auch auf agile Software-Entwicklungsteams in vielen Unternehmen zu. Eine zu hohe Arbeitsbelastung kann zu diversen negativen Effekten, wie einem erhöhten Krankenstand, dem Wohlbefinden der Mitarbeitenden oder reduzierter Produktivität führen. Es ist zudem bekannt, dass sich die Arbeitsbelastung in der Wissensarbeit auf die Qualität der Arbeitsergebnisse auswirkt. Dieser Forschungsbeitrag identifiziert potentielle Faktoren der Arbeitsbelastung der Mitglieder eines agilen Software-Entwicklungsteam bei der Otto GmbH & Co KG. Auf der Grundlage der Faktoren präsentieren wir Maßnahmen zur Reduzierung von Arbeitsbelastung und erläutern unsere Erkenntnisse, die wir im Rahmen eines Experiments validiert haben. Unsere Ergebnisse zeigen, dass bereits kleinteilige Maßnahmen, wie das Einführen von Ruhearbeitsphasen während des Arbeitstages, zu positiven Effekten bspw. hinsichtlich einer gesteigerten Konzentrationsfähigkeit führen und wie sich diese auf die Qualität der Arbeitsergebnisse auswirken.

Schlüsselwörter: Agile Software Entwicklung, Arbeitsbelastung, Arbeitsqualität, Remote Work

**Abstract:** Due to the Covid 19 pandemic and the associated effects on the world of work, the burden on employees has been brought into focus. This fact also applies to agile software development teams in many companies due to the extensive switch to remote work. Too high a workload can lead to various negative effects, such as increased sick leave, the well-being of employees, or reduced productivity. It is also known that the workload in knowledge work impacts the quality of the work results. This research article identifies potential factors of the workload of the agile software development team members at Otto GmbH & Co KG. Based on the factors, we present measures to reduce workload and explain our findings, which we have validated in an experiment. Our results show that even small-scale actions, such as the introduction of rest work phases during the working day, lead to positive effects, for example, increased ability to concentrate and how these affect the quality of the work results.

## 1. Einleitung

Die Arbeitsbelastung ist durch die Covid 19 Pandemie in den Fokus vieler Unternehmen gerückt. Das Arbeiten in sitzender Haltung (Metz und Rothe 2017, S.27) oder die ständige Kommunikation und Erreichbarkeit über digitale Plattformen und Messenger (Syles 2010, S. 389 f.) sind nur einige der Faktoren, die im Umfeld der Softwareentwicklung zu einer hohen Arbeitsbelastung führen können (Mojtahedzadeh et al. 2021).

Der Grad der Arbeitsbelastung kann Effekte auf die Leistungsfähigkeit von Arbeitnehmenden ausüben und das Wohlbefinden beeinträchtigen. Diese Effekte führen zudem zu Auswirkungen in Unternehmen. Hierzu gehören erhöhte Fehlzeiten, reduzierte Produktivität, vermehrte Fehler und ein schlechteres Betriebsklima (Metz und Rother 2017), was insgesamt die Arbeitsqualität negativ beeinflussen kann. Eine hohe Arbeitsqualität ist jedoch entscheidend für die Produktqualität und beeinflusst darüber auch die Kundenzufriedenheit. Die erforderliche Arbeitsqualität muss gegeben sein, um Kundenanforderungen erfüllen zu können und so die Kundenbeziehung zu festigen. Daher ist eine hohe Arbeitsqualität erstrebenswert. Bisherige Studien haben sich unter anderem mit dem Einfluss der Workload auf die Performance bei standardisierten Aufgaben (Fan und Smith 2017) beschäftigt oder verschiedene Einflussfaktoren auf die Softwarequalität (Acuña et al. 2009) untersucht.

Es stellt sich daher die Frage, ob sich Arbeitsbelastung und Arbeitsqualität tatsächlich beeinflussen und verringerte Arbeitsbelastung zu höherer Arbeitsqualität führt. Diese Fragestellung haben wir in einer Fallstudie untersucht und präsentieren in diesem Artikel die Ergebnisse.

Dieser Artikel ist wie folgt aufgebaut: Zunächst erläutern wir die Grundlagen zur Arbeitsbelastung und die Auswirkungen als Einflussfaktor für die Arbeitsqualität. Im Folgenden stellen wir unsere Forschungsmethodik dar, bevor wir die Ergebnisse der Studie diskutieren und den Artikel mit einem Fazit abschließen.

## 2. Theoretische Grundlagen

Arbeitsbelastung wird in der Norm DIN ES ISO 6385:2016-12 als die *„Gesamtheit der äußeren Bedingungen und Anforderungen im Arbeitssystem, die auf den physiologischen und/oder psychologischen Zustand einer Person einwirken"* (Hasselhorn und Müller 2004, S. 18) definiert. Da es sich bei Tätigkeiten im Umfeld der agilen Softwareentwicklung um Wissensarbeit handelt, stehen dort vor allem die psychischen Belastungsfaktoren im Vordergrund. Einflüsse, die von außen auf die Psyche einer Person einwirken, können sowohl aus dem Arbeitsinhalt, den Ausführungsbedingungen, der Arbeitsorganisation als auch aus den sozialen Beziehungen resultieren (Metz und Rothe, S. 7). Die Auswirkungen dieser Faktoren auf den Arbeitenden sind psychische Beanspruchungen (Metz und Rothe, S. 17). Die Stärke der Beanspruchung einer Person hängt dabei von ihren individuellen Bewältigungsfähigkeiten ab (Metz und Rothe, S. 11f.). Eine kurzfristige Folge von psychischer Beanspruchung, die im Laufe eines Arbeitstages auftritt, kann unter anderem psychische Ermüdung sein (Metz und Rothe, S. 13). Diese kann die Informationsverarbeitung des Arbeitenden beeinträchtigen und zu Leistungsschwankungen, verminderter Konzentration, Fehlhandlungen und Erleben von Müdigkeit führen (Metz und Rothe, S. 13). Der Einfluss der Workload auf die Performance bei der Ausführung von verschiedenen Aufgaben, wie Reaktionstests oder visuellen Suchaufgaben, wurde in der Literatur bereits ausführlich behandelt. Die Ergebnisse von Fan und Smith (2017) zeigen, dass eine erhöhte Workload mit einer erhöhten Müdigkeit der Teilnehmenden einhergeht, die wiederum einen negativen Einfluss auf die Performance hatte. Ein direkter Zusammenhang zwischen der Workload und der Performance konnte nicht festgestellt werden. Die Ergebnisse der Studie decken sich folglich mit der Theorie von Belastung, Beanspruchung und deren Folgen.

Der wesentliche Bestandteil der Ergebnisse in der Softwareentwicklung ist das auslieferbare Artefakt. Entscheidend für den Erfolg dieser Artefakte ist deren Qualität. Der Einfluss verschiedener Faktoren auf die Softwarequalität war in der Vergangenheit bereits Gegenstand von Veröffentlichungen. Acuña et al. untersuchten den Einfluss von Persönlichkeit, Team-Prozessen und Aufgabencharakteristiken auf die Softwarequalität und die Jobzufriedenheit. Dabei verwendeten sie verschiedene Qualitätskriterien, wie die Anzahl und Verknüpfung von Modulen (Modularisierung), die Anzahl der durch Tests erkannten Fehlern (Testbarkeit), die Anzahl an erfüllten Anforderungen (Funktionalität) und die Anzahl an wiederverwendeten Modulen (Wiederverwendbarkeit) (Acuña et al. 2009). Die Studie von Bettenburg und Hassan (2012) betrachtete den Einfluss von sozialen Interaktionen und Kollaboration auf Versionsverwaltungsplattformen während des Entwicklungsprozesses auf die Anzahl an gemeldeten Fehlern in großen Open Source Softwareprojekten.

## 3. Methodisches Vorgehen

In diesem Abschnitt werden das deduktive Forschungsdesign und das dazugehörige methodische Vorgehen dieser wissenschaftlichen Arbeit aufgezeigt (siehe Abbildung 1).

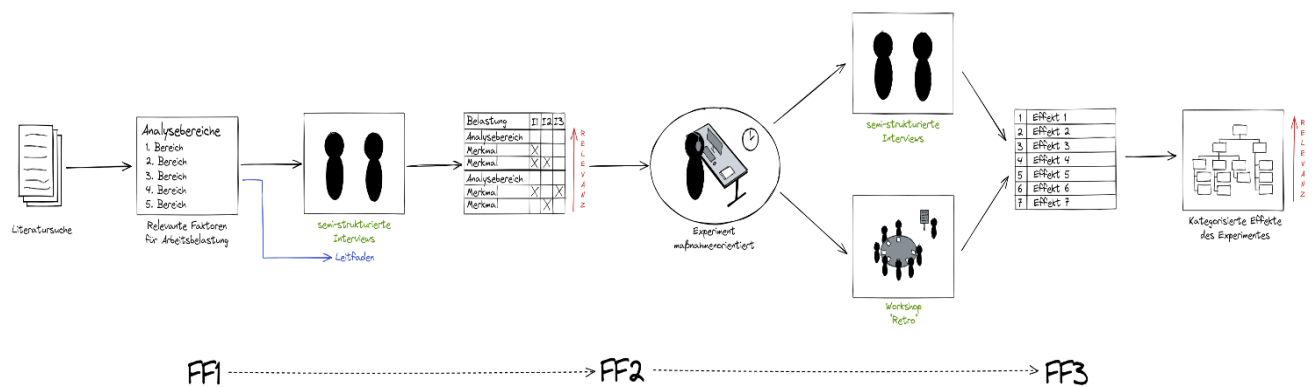

Abbildung 1: Methodisches Design

Mit Hilfe der Goal-Question-Metric-Plus-Methode[1], einer Methode zur systematischen Strukturierung von übergeordneten Zielen, Teilzielen, daraus abgeleiteten Unterfragen und eigens aufgestellten Metriken zur Beantwortung der Fragen (Basili et al. 2007), wurde die übergeordnete Forschungsfrage in sich daraus abzuleitende Forschungsfragen unterteilt, welche wie folgt definiert wurden:

*FF1: Welche Faktoren von Arbeitsbelastung liegen vor?*

Während der Recherche wurden mehrere Faktoren für Arbeitsbelastung definiert. Wir möchten mit dieser Forschungsfrage klären, welche der möglichen Faktoren im Software-Entwicklungsteam bei Otto für Arbeitsbelastung sorgen.

*FF2: Wie lassen sich die vorliegenden Faktoren von Arbeitsbelastung reduzieren?*

Diese Frage impliziert auch die Frage, ob sich die Arbeitsbelastung verursachenden Faktoren per se reduzieren lassen. Auf Basis dieser Forschungsfrage entwickeln wir das Experiment.

*FF3: Wie wirkt sich die Reduzierung der Arbeitsbelastung auf die Arbeitsqualität aus?*

Unsere dritte Forschungsfrage zielt darauf ab zu analysieren, inwiefern eine reduzierte Arbeitsbelastung eine Auswirkung auf die Arbeitsqualität des agilen Software-Entwicklungsteams hat.

Zur Beurteilung der These, dass eine reduzierte Arbeitsbelastung zu erhöhter Arbeitsqualität führt, wurden zunächst aus der Literatur Definitionen und Faktoren von Arbeitsbelastung und Arbeitsqualität entnommen. Das zur Beantwortung der Forschungsfrage entwickelte Experiment wurde in einem agilen Softwareentwicklungsteam der E-Commerce-Abteilung der Otto GmbH & Co KG, eines in Deutschland ansässigen, international agierenden Einzelhandelskonzerns, durchgeführt. Durch die Covid-19-Pandemie arbeitete das Team remote, weshalb auch sämtliche Interaktionen, wie z.B. die Datenerhebung, remote erfolgten.

---

[1] Anhang A: GQM+ Dokumentation.

Die qualitative Datenerhebung fand von April bis Juli 2021 statt. Vor Beginn des Experiments wurde ein Workshop zur Schaffung eines Überblicks über mögliche Bereiche von Arbeitsbelastung mit einem ersten Team durchgeführt. In diesem wurde jedoch genannt, dass das Team in keinem der Bereiche Arbeitsbelastung verspüre. Daraufhin erfolgte ein Wechsel des Teams. Ebenfalls fand ein methodischer Wechsel statt, indem vier Mitglieder unmittelbar in einem semi-strukturierten Interview befragt wurden, um gezielter nach möglichen belastenden Eigenschaften ihrer Arbeit Fragen zu können. Der Interviewleitfaden[2] orientierte sich an den Fragen des Fragebogens „Screening psychischer Arbeitsbelastung" (SPA) von Metz/Rothe 2017. Nach dem zweiwöchigen Experiment wurden dieselben Personen erneut mittels semi-strukturiertem Interview[3] befragt. Zusätzlich fand ein Retrospektiven-Workshop[4] statt. Dieser sollte auch nicht interviewten Teammitgliedern einen Anreiz zur aktiven Teilnahme am Experiment bieten, damit sie ebenfalls über die gesammelten Erfahrungen berichten können. Da das Team keine der in vergleichbaren Studien verwendeten Metriken zu Bewertung von Softwarequalität einsetzt, musste man sich im Weiteren auf die subjektive Wahrnehmung der Arbeitsqualität der Teammitglieder stützen. Die erhobenen Daten wurden mithilfe einer Inhaltsanalyse anhand der Analysebereiche von Metz und Rothe (2017) aus dem Screening psychischer Arbeitsbelastung ausgewertet. Für die Auswertung der Interviewreihe nach Durchführung der Maßnahmen wurden übergeordnete Kategorien gebildet, zu denen die beobachteten Effekte zugeordnet worden.

### 4. Analyse und Diskussion
#### 4.1 Feststellung der Arbeitsbelastung

In Tabelle 1 werden die Ergebnisse der ersten vier Interviews dargestellt.[5] Die Ergebnisse sind aufgeschlüsselt nach den Bereichen des Screenings psychischer Arbeitsbelastung von Metz und Rothe (2017, S. 47 ff.) und den einzelnen Interviewpersonen (I1, I2, I3, I4) zugeordnet. Ein „X" symbolisiert eine festgestellte Arbeitsbelastung in diesem Bereich.

| ID | Arbeitsbelastungsbereich | I1 | I2 | I3 | I4 |
|---|---|---|---|---|---|
| *2* | *Entscheidungsspielraum* | | | | |
| 2.1 | Selbstbestimmte Tätigkeiten und ihre Relevanz innerhalb der Abteilung | | X | | |
| 2.2 | Planung und Freiheit bei der Durchführung der Arbeit | | | | |
| 2.3 | Überstunden und Auswirkung auf die Work-Life-Balance | X | X | | X |
| *3* | *Komplexität und Variabilität* | | | | |
| 3.1 | Art der Tätigkeiten | | | | |
| 3.2 | Abwechslung der Aufgaben und ihrer Bearbeitung | | | | |
| 3.3 | Wiederholende Sinnesleistungen | X | X | X | X |
| 3.4 | Zusammenarbeit mit Teammitgliedern | | | | |
| *4* | *Qualifikationserfordernisse* | | | | |
| 4.1 | Einsatz von gelernten Kompetenzen und Fähigkeiten | | | | |
| 4.2 | Erforderliche Weiterbildungen | | | | |
| *5* | *Risikobehaftete Arbeitssituationen* | | | | |
| 5.1 | Parallele Aufgabenbearbeitung | X | X | X | X |

---

[2] Anhang B: Interviewleitfaden 1: Feststellung der Arbeitsbelastung.
[3] Anhang C: Interviewleitfaden 2: Auswertung der Maßnahme.
[4] Anhang D: Retrospektiven-Workshop.
[5] Anhang E: I 1 bis I4 – Protokolle der Interviews zu Leitfaden 1.

| 5.2 | Auswirkung schwerwiegender Fehler | | | | |
|---|---|---|---|---|---|
| 5.3 | Zeitnot | | X | | X |
| 5.4 | Abhängigkeit der Teammitglieder von der IP | | | | X |
| *6* | *Belastende Ausführungsbedingungen* | | | | |
| 6.2 | Eigene Fehlerbehebung | | | | |
| 6.3 | Arbeitsunterbrechung und Auswirkung auf die Konzentration | X | | X | X |
| 6.4 | Verfügbarkeit und Beschaffungsaufwand benötigter Informationen | | X | X | X |
| 6.5 | Unterstützungskultur im Team | | | | |
| 6.6 | Häufigkeit von persönlichen Gesprächen | | | | |
| *7* | *Soziale Beziehungen* | | | | |
| 7.1+7.2 | Teamzusammenhalt | | | | |
| *9* | *Körperliche Beanspruchung* | | | | |
| 9.1 | Körperliche Belastungssymptome | X | X | X | |

Tabelle 1: Festgestellte Belastungen

Wie Tabelle 1 zu entnehmen ist, wurde in den Bereichen Soziale Beziehungen und Qualifikationserfordernisse keine Arbeitsbelastung festgestellt. I2 gab unter Frage 2.1 an, dass die Art der Tätigkeit und die Themen meist gesetzt seien und sie darauf wenig Einfluss habe. Dadurch liegt eine Arbeitsbelastung in diesem Bereich vor. Des Weiteren wurde festgestellt, dass drei Interviewpersonen (I1, I2, I4) regelmäßig Überstunden machen. Dies stellt eine Belastung dar, I1 gab an „nach einem langen Tag ist man kaputter" (I1, Frage 2.3). Es wurde jedoch von den drei Interviewpersonen angeben, Überstunden stets freiwillig und nach eigenem Ermessen zu machen.

Im Bereich Komplexität und Variabilität wurde festgestellt, dass alle Interviewpersonen regelmäßig sich wiederholende Sinnesleistungen durchführen müssen. Dies liegt vor allem an der Art der Tätigkeiten. I1 und I3, welche in der Softwareentwicklung tätig sind, müssen bspw. häufig Dateneingaben im Rahmen der Programmierung leisten.

Alle Interviewpersonen gaben bei 5.1 an, regelmäßig parallel Aufgaben zu bearbeiten, I4 bearbeite parallel zu Meetings andere Aufgaben. Außerdem wurde bei einer Interviewperson (I1) zu 5.2 festgestellt, dass Fehler schwerwiegend sein können, diese jedoch aufgrund des Systems verhindert würden. Ebenfalls wurde Arbeitsbelastung durch Zeitnot (I2, I4) identifiziert. Diese entstehe besonders dann, wenn keine Pause zwischen Meetings vorhanden sei (I2) oder das persönlich gesetzte Arbeitsziel zu hoch sei (I4). Unter Frage 5.4 gab I4 an, dass Teammitglieder von ihrer Zuarbeit abhängig seien, die anderen Interviewpersonen schätzten die Abhängigkeit für sehr gering ein.

Im Bereich belastende Ausführungsbedingungen gaben bei Frage 6.3 drei Interviewpersonen (I1, I3, I4) an, regelmäßig unterbrochen zu werden. Die Unterbrechung bezog sich meist auf irrelevante Benachrichtigungen durch Teams. I2, bei der keine Belastung durch Arbeitsunterbrechungen festgestellt wurde, gab an, ihre Benachrichtigungseinstellungen bereits angepasst zu haben. Die Verfügbarkeit von benötigten Informationen und ein hoher Beschaffungsaufwand wurde von drei Interviewpersonen (I2, I3, I4) bemängelt. Dokumentationen seien meist nicht vorhanden (I1, I2, I3, I4). Der Einsatz von Dokumentationen wurde jedoch von den interviewten Personen im Bereich der Softwareentwicklung in Frage gestellt, da diese auch Pflege benötigen würden (I1) und es auch ohne Dokumentationen funktionieren würde (I3).

Drei Interviewpersonen (I1, I3, I4) gaben bei Frage 9.1 an unter körperlichen Belastungssymptomen zu leiden. Es wurde von Rückenschmerzen, Schulterschmerzen und Kopfschmerzen berichtet.

## 4.2 Ableitung der Maßnahmen zur reduzierten Arbeitsbelastung

Auf Basis der festgestellten Arbeitsbelastung wurde mithilfe von Literaturrecherche und Brainstorming ein Maßnahmenpaket zur Senkung der vorhandenen Arbeitsbelastung zusammengestellt.

Als Maßnahme gegen allgemeine Beanspruchung können regelmäßige und bewusste Pausen helfen. Um auch die festgestellte Belastung durch Rückenschmerzen zu verringern, könnten die Pausen aktiv verbracht werden, z.B. Durchführung kurzer Yoga-Videos. So soll es sinnvoll sein, alle 40 Minuten eine aktive Pause zu nehmen, um die durch Sitzhaltung beanspruchte Muskulatur zu entspannen (Ding et al. 2020). Zudem sei es laut der Bundesanstalt für Arbeitsschutz wichtig, dass Pausen geplant und vorhersehbar genommen werden und gleichmäßig auf den Arbeitstag verteilt sind (Bundesanstalt für Arbeitsschutz und Arbeitsmedizin 2019).

Wie bei der agilen Praktik Energized Work aus Extreme Programming sollten Überstunden möglichst vermieden werden, um einen Ausgleich zwischen Arbeit und Freizeit sicherzustellen und energievolle Arbeit zu ermöglichen (Beck and Andres 2004). Hierfür könnte die Vorgabe helfen, dass Überstunden der einen Arbeitswoche in der darauffolgenden Woche ausgeglichen werden müssen.

Zur Verringerung des Beschaffungsaufwandes für benötigte Informationen können Ansprechpersonen für bestimmte Themen festgehalten werden. Bei Fragen kann sich direkt an die zuständige Person gewandt werden, ohne dass vorab Zeit zur Feststellung dieser investiert werden muss. Außerdem könnte es hilfreich sein die Teammitglieder anzuhalten, Dokumentationen zu erstellen und zu pflegen. Dadurch entsteht jedoch auch ein zusätzlicher Zeitaufwand.

Die Einführung von Ruhearbeitsslots kann Arbeitsunterbrechung während vorgegebener Zeiten vermeiden. Darüber hinaus kann die Anpassung der Benachrichtigungseinstellungen von Teams und Outlook helfen, die Unterbrechungen durch Benachrichtigungen zu verhindern, wie auch die Studie von Sykes (2010) zeigt. Im Rahmen des Experimentes wurde sich dazu entschieden, Ruhearbeitsslots beim zugewiesenen Team einzuführen. Über einen Zeitraum von zwei Wochen sollten die Teammitglieder zwei einstündige Ruhearbeitsslots pro Arbeitstag nehmen. Die Slots sollten sich vorab, für die Kollegen sichtbar, in den Kalender eingetragen werden. Um Ablenkungen durch Teams und Outlook zu vermeiden, sollten diese während der Ruhearbeitsslots geschlossen werden. Falls Teams oder Outlook während des Ruhearbeitsslots zwingend benötigt werden, sollten die Benachrichtigungen ausgestellt werden, um die Ablenkung zumindest zu verringern. Es wurde sich für die Ruhearbeitsslots entschieden, da drei Interviewpersonen über Arbeitsunterbrechungen, insbesondere durch Teams, geklagt hatten. Zudem können unterbrochene Personen bis zu 15 Minuten benötigen, um wieder in ihre vorherige Aufgabe zurückzufinden (Jackson et al. 2001, nach Stich 2020), weshalb durch die Maßnahme eine Zeitersparnis bei konzentrierter Arbeit stattfinden kann. Ebenfalls können die Auswirkungen dieser Maßnahme durch konkretes Erfragen der Veränderungen während der Slots, eindeutig festgestellt werden. Dadurch wurde sich erhofft, trotz der kurzen Dauer des Experiments bereits Veränderungen zu erzielen. Des Weiteren erklärte sich auch die Projektleitung mit dieser Maßnahme einverstanden.

### 4.3 Auswirkungen der reduzierten Arbeitsbelastung auf die Arbeitsqualität

Die Ergebnisse des Workshops[6] (W) und der Interviews[7] nach der Durchführung des Experiments wurden in drei Bereiche aufgeteilt (siehe Tabelle 2). Dies sind zum einen die positiven und negativen Effekte, die die teilnehmenden Personen während des Experimentes beobachten konnten und zum anderen die Gründe, warum die Ruhearbeitsslots nicht eingehalten wurden.

---

[6] Anhang D: Retrospektiven-Workshop.
[7] Anhang F: I 1 bis I4 – Protokolle der Interviews zu Leitfaden 2.

| Ebene 1 | Ebene 2 | Ebene 3 | Verweise |
|---|---|---|---|
| Positive Effekte | Konzentration | Bewusster Fokus auf ein Thema | I1, I2, I3, I4, W |
| | | Bewusster Zeit nehmen | I2, I3, I4 |
| | | Gründlichere Arbeit | I4 |
| | Entspannter Arbeiten | Ruhearbeitslots sind bekannt | I2, I3 |
| | | Weniger Termine | I1, W |
| | Bewussteres Arbeitsverhalten/-umfeld | Ergonomische Sitzhaltung | I3 |
| | | Reflexion des Arbeitstages | I3, I4 |
| | | Anpassung der Benachrichtigungen | I3 |
| | | Teamgespräche über Fokussierung | I3 |
| Negative Effekte | Kommunikation | Asynchrone, verlängerte Kommunikation | I3, W |
| | | Erhöhter Aufwand nach Slots | I3, W |
| | | Blockieren der Arbeit von Mitarbeitenden | I1 |
| Gründe für die Nichteinhaltung | Art der Aufgabe | Informationen benötigt | I1, I2, I4, W |
| | | Daily Action Pair | I3 |
| | Ablauf des Arbeitstages | Arbeitszeiten | I1, W |
| | | Meetings | I2, I4, W |
| | Kommunikation | Angst etwas zu verpassen | W |
| | Inkonsequenz | Vergessene Ruhearbeitsslots | I1, W |

Tabelle 2: Auswertung der Ruhearbeitsslots

Es wurde genannt, dass für die Durchführung der aktuellen Aufgabe Informationen von Kollegen benötigt wurden, wodurch die Programme nicht vollständig geschlossen werden konnten, da eine Kommunikation zwingend nötig war (I1, I2, I4, W). Ebenfalls spielte der Ablauf des Arbeitstages eine große Rolle bei der Umsetzbarkeit der Maßnahmen. Das Team hatte sich zu Beginn des Experimentes auf zwei feste Slots (9-10 Uhr und 13-14 Uhr) geeinigt, während denen die Ruhearbeit durchgeführt werden kann. Diese Slots passten jedoch nicht immer zu den Arbeitszeiten aller Teammitglieder, weshalb die Slots nicht wahrgenommen wurden (I1, W). Eine individuelle Durchführung zu einer anderen Zeit fand nur selten statt (W). Ein weiterer wichtiger Punkt für das Ausfallen von Ruhearbeitsslots waren kollidierende Meetings (I2, I4, W). Neben bereits vor dem Experiment geplanter Meetings hat eine interviewte Person ebenfalls aufgrund von persönlicher Präferenz und der Rolle als Produktmanager auf Slots am Nachmittag verzichtet (I2). Des Weiteren wurde die Angst etwas zu verpassen als Grund für die Nichtwahrnehmung der Ruhearbeitsslots genannt (W).

Als negativer Effekt wurde unter anderem eine asynchrone und dadurch zeitlich verlängerte Kommunikation genannt (I3, W). Zudem mussten die Nachrichten, die sich während des Ruhearbeitsslots angesammelt haben, im Anschluss auf einmal beantwortet werden. Dies wurde teilweise als erhöhter Aufwand wahrgenommen (I3, W). Ein weiterer negativer Aspekt war das Blockieren der Arbeit von Kollegen, die für die Durchführung ihrer Aufgabe Informationen von einem benötigten (I1). Die Aussage: „Ich werde durch die Benachrichtigungen zwar rausgerissen, aber dafür kann der andere dann weiterarbeiten" (I1) zeigt, dass die Person die Unterstützung von Mitarbeitenden als wichtiger ansieht als die Konzentration auf die eigene Aufgabe.

Die positiven Effekte wurden in drei Kategorien zusammengefasst: Konzentration, Entspannter Arbeiten und Bewussteres Arbeitsverhalten/-umfeld. Alle interviewten Personen berichteten von einer höheren Konzentration durch den Fokus auf ein Thema oder eine Aufgabe, die sie sich für die Zeit der Ruhearbeitsslots vorgenommen hatten (I1, I2, I3, I4). Dies führte dazu, dass sich zu gewissen Themen ausführlicher Gedanken gemacht und sich mehr Zeit für Aufgaben genommen wurde, als dies im normalen Arbeitsalltag der Fall sei (I2, I3, I4). Darauf aufbauend wurde ebenfalls genannt, dass dadurch Ergebnisse weniger hektisch mit Kollegen geteilt worden seien. Stattdessen wurde sich die Zeit genommen, die Ergebnisse erst nach einer gründlicheren Betrachtung herauszugeben, was zu weniger Korrekturen geführt habe (I4). Ein entspannteres Arbeiten wurde durch zwei Faktoren ermöglicht. Zum einen wurde der Druck ständig erreichbar zu sein dadurch genommen, dass die Ruheslots im Team bekannt waren und daher die Kollegen damit rechneten, zu bestimmten Zeiten nicht unmittelbar eine Antwort zu erhalten (I2, I3). Zum anderen wurden die durch die Slots entstandenen terminfreien Zeiten als positiv wahrgenommen. Dies war ebenfalls bei Personen der Fall, die die Kommunikationstools während dieser Zeit nicht schließen konnten (I1). Einige interviewte Personen nannten ebenfalls, dass sie während der Ruhearbeitsslots auf eine ergonomische Sitzhaltung geachtet (I3) oder den eigenen Arbeitstag reflektiert haben (I3, I4). Durch das Home-Office sei man dazu gezwungen gewesen über Teams zu kommunizieren (I3) und habe „eher ignoriert, dass da ständig was aufflackert" (I3). Das Experiment wurde daher als Anlass genommen die Benachrichtigungen zu managen. Zudem fanden Teamgespräche über Fokussierung statt (I3).

Aufgrund der Aussagen in den Interviews und dem Workshop kann man davon ausgehen, dass die Arbeitsbelastung zu einem gewissen Grad gesenkt wurde. Die Unterbrechungen durch Kollegen wurden während der Slots minimiert, weshalb die Person in diesem Zeitraum weniger beansprucht wurde. Durch den Fokus auf eine Aufgabe konnte außerdem die parallele Aufgabenbearbeitung als reduziert bezeichnet werden. Zudem führte das Experiment zu einer Verbesserung des individuellen Bewältigungsverhaltens indem auf ein bewussteres Arbeitsverhalten und -umfeld geachtet wurde.

Da keine objektiven Metriken zur Bewertung der Arbeitsqualität zur Verfügung standen, konnten die interviewten Personen lediglich nach einer Einschätzung gefragt werden. Zwei Teilnehmende konnten dazu aufgrund der Kürze des Experimentes keine Aussage treffen (I1, I2). Die beiden anderen bewerteten die Auswirkungen des Experimentes auf die Arbeitsqualität als positiv (I3, I4). Diese Einschätzungen lassen sich durch die identifizierten positiven Effekte stützen. Durch die höhere Konzentration während der Ruhearbeitsslots, einem besseren Fokus auf das zu bearbeitende Thema und auch den gründlicheren Abschluss von Aufgaben, kann man auf eine höhere Qualität der erarbeiteten Ergebnisse schließen. Dies deckt sich ebenfalls mit den Erkenntnissen der Studie von Foroughi et al. (2014). Dort zeigen sie, dass Unterbrechungen zu einer allgemein verringerten Qualität bei kreativen Schreibaufgaben führen (Foroughi et al. 2014).

## 5. Limitierungen

Trotz eines methodischen Vorgehens sind einige Limitationen während des Forschungsprojektes zu berücksichtigen. Der Zeitraum, in dem das Experiment durchgeführt worden ist, kann als zu kurz betrachtet werden, um eindeutige Ergebnisse festzustellen. Dies wird dadurch verstärkt, dass die Auswertung des Experiments eine inkonsequente Umsetzung der Ruhearbeitsslots zeigte. Ein Grund hierfür ist unter anderem, dass die Umsetzbarkeit der Ruhearbeitsslots stark abhängig von der Rolle der Person ist. Für Rollen, deren Aufgaben häufig aus Absprachen mit Kollegen besteht, stellt das Wahrnehmen der Slots im Arbeitsalltag beispielsweise eine größere Hürde dar.

Die Messung der Arbeitsqualität konnte nicht durch objektive Messdaten erfolgen. Hierfür konnte man sich nur auf die subjektiv wahrgenommene Arbeitsqualität berufen, welche jedoch, dadurch dass sie lediglich subjektiv ist, beeinflussbarer und weniger konsistent ist, als wenn sie anhand von Daten messbar gewesen wäre (siehe Kapitel 4.3). Eine weitere Limitierung stellt dar, dass die Beanspruchung und Auswirkung einzelner Belastungsfaktoren auf die gesamte psychische Beanspruchung individuell ist. So können Belastungsfaktoren, die für die eine Person eine entscheidende Rolle für die Performance und die Arbeitsqualität spielen, für eine andere Person einen untergeordneten Einfluss haben.

## 6. Fazit

Der Grad der Arbeitsbelastung seiner Beschäftigten ist für ein Unternehmen von großer Bedeutung, da er sich sowohl auf das Wohlbefinden als auch die Arbeitsergebnisse auswirkt. Daher sollte Ziel jedes Unternehmens sein, die Belastung der Mitarbeitenden so gut wie möglich zu reduzieren. Mit Hilfe des ersten Interviews wurden die belastenden Faktoren des betrachteten Teams identifiziert. Neben Überstunden und mangelnder Informationsverfügbarkeit wurden dabei vor allem Belastungen im Bereich der parallelen Aufgabenbearbeitung und häufige Unterbrechungen festgestellt.

Anhand der ermittelten Belastungsfaktoren wurde eine geeignete Maßnahme abgeleitet, um die Arbeitsbelastung der Mitarbeitenden zu verringern. Mit Hilfe von Ruhearbeitsslots konnten die Unterbrechungen während zwei Stunden am Tag eliminiert werden. Auch die parallele Arbeit konnte in einigen Fällen reduziert werden, da sich in terminfreien Zeiten nur auf eine Aufgabe konzentriert werden konnte. Das Anregen eines bewussteren Arbeitsverhaltens und -umfelds ist zudem ein Nebeneffekt, der weitere Belastungsfaktoren reduzieren kann. Mittels des Workshops und der zweiten Interviews nach der Durchführung der Maßnahme wurden die Effekte auf die Arbeit der Teilnehmenden untersucht (siehe Kapitel 4.3). Dabei wurde eine höhere Konzentration auf die aktuelle Aufgabe und ein entspannteres Arbeiten festgestellt. Daraus kann man ebenfalls auf eine höhere Arbeitsqualität schließen, was in zwei der Interviews bestätigt wurde. Die Überlegung, dass eine Reduzierung der Arbeitsbelastung zu einer höheren Arbeitsqualität führt, lässt sich daher, unter Berücksichtigung des Kontexts und der Limitierungen unserer Arbeit, bestätigen. Wir empfehlen die Erkenntnisse unserer Arbeit auf andere Forschungskontexte zu überführen, um zu überprüfen, ob diese Effekte auch bei anderen agilen Software-Entwicklungsteams in anderen Branchen und Unternehmen beobachtet werden können.

## Anhänge

Die folgenden Anhänge sind unter folgendem Link in der AcademicCloud abrufbar:
https://sync.academiccloud.de/index.php/s/yGHv1QJFZMzqI92

Anhang A – GQM+ Metric
Anhang B – Interviewleitfaden 1: Feststellung der Arbeitsbelastung
Anhang C – Interviewleitfaden 2: Auswertung der Maßnahme
Anhang D – Retrospektiven Workshop
Anhang E – I 1 bis I4 – Protokolle der Interviews zu Leitfaden 1
Anhang F – I 1 bis I4 – Protokolle der Interviews zu Leitfaden 2

## Quellenverzeichnis